\documentclass[floatfix,aps,pra,twocolumn,showpacs,superscriptaddress,10pt]{revtex4-2}

\usepackage[utf8]{inputenc}
\usepackage[english]{babel}
\usepackage{caption}
\usepackage{mathtools}
\usepackage{float}
\usepackage{siunitx}
\usepackage{amsmath}
\usepackage{subcaption}
\usepackage{amsfonts}
\usepackage{amssymb}
\usepackage{graphicx}
\usepackage{color}
\usepackage[colorlinks=true,citecolor=blue,linkcolor=blue,urlcolor=blue]{hyperref}
\usepackage{times}
\usepackage{amstext}
\usepackage{latexsym}
\usepackage[running,mathlines]{lineno}
\usepackage{verbatim}
\usepackage{physics}
\usepackage{bm}
\usepackage{ulem}
\usepackage{qcircuit}
\usepackage[version=3]{mhchem}
\usepackage{lipsum}
\usepackage{ragged2e}

\newcommand{\beq}{\begin{equation}}
\newcommand{\eeq}{\end{equation}}
\newcommand{\barr}{\begin{eqnarray}}
\newcommand{\earr}{\end{eqnarray}}

\usepackage{orcidlink}
\newcommand{\UFSCar}{Departamento de Física, Universidade Federal de São Carlos, Rodovia Washington Luís, km 235 - SP-310, 13565-905 São Carlos, SP, Brazil}

\begin{document}
\raggedbottom



\title{Interference Between Electromagnetic and Mechanical Waves}

\author{Alexandre Cesar Ricardo~\orcidlink{0000-0003-1282-9484}}\email{alexandre.ricardo@df.ufscar.br}
\affiliation{\UFSCar}

\author{Ciro Micheletti~Diniz~\orcidlink{0000-0002-7602-0468}}\email{ciromd@outlook.com.br}
\affiliation{\UFSCar}

\author{Celso Jorge Villas-Bôas~\orcidlink{0000-0001-5622-786X}}\email{celsovb@df.ufscar.br}
\affiliation{\UFSCar}

\begin{abstract}


Classically, wave interference is a phenomenon that can be explained by considering only the waves themselves, that is, without the need to consider the apparatus that monitors or observes them. Thus, in classical theories, interference can only occur between waves of the same nature. In quantum theory, the observed results require a description of the system and its measuring apparatus, which allows us to rethink the explanation of various natural phenomena. In this paper, we consider the ion-trap platform to study the interference of waves with different physical natures, specifically the electromagnetic and mechanical. At first, we drive two lasers onto a single-trapped ion to produce Jaynes-Cummings and Carrier interactions, where we verify that, depending on the phase relationship between the coherent state of the vibrational (mechanical) mode and the Carrier pulse (electromagnetic wave), the interactions enhance or cancel out population transfer to the electronic state of the ion, that works out as our measuring apparatus for those waves. 
Extending our result to an ion chain, we verify that a precise modulation of the Carrier Rabi frequency and phase (electromagnetic pulse) according to the amplitude of the incoming mechanical coherent state in the ion chain enables creating either constructive or destructive interference with propagating pulses, in which the electronic state of the driven ion is, respectively, populated more and faster, or transparent to both pulse waves, when the information flux behaves as if no external fields are applied. Finally, this new type of controlled interference between waves of different natures allows us to propose new hybrid quantum devices, such as transistors or filters of wave packets, where photonic (phononic) pulses control the passage of phononic (photonic) waves.


\end{abstract}

\maketitle

\section{Introduction}

Wave phenomena are fundamental to much of classical physics, offering a core framework to describe disturbances traveling through various media like electromagnetic fields, elastic solids, and fluids~\cite{freegarde2012introduction,dorfler2020mathematics,towne1967wave}. In the past, waves were seen as collective oscillations that could interfere, diffract, and refract, described mathematically by continuous fields~\cite{hecht2017optics, born1959principles}.  Interestingly, classical wave theory naturally appears as a limiting case within quantum mechanics, allowing quantum descriptions to recover familiar classical results in limiting regimes~\cite{Bohr01071913, zurek2003decoherencetransitionquantumclassical}. The development of quantum mechanics in the early 20th century introduced the wave-particle duality, which showed that light, represented by photons, and matter itself, described as matter waves or phonons in quantized vibrational states, can display wave-like behavior~\cite{DEBROGLIE1923,  debroglie1925,10.1063/1.3047826}. More recently, this shared feature has been exploited in a particle description of interference~\cite{PhysRevLett.134.133603,diniz2025pulsedlasercontinuousparticle}, which was traditionally explained by waves. 

In classical physics, interference is a physical phenomenon that arises when two or more coherent waves of the same nature overlap at a given point in space and time, yielding a new wave pattern~\cite{hecht2017optics, born1959principles}. This superposition modifies the resultant amplitude according to the linear superposition principle and yields a new wave pattern. The resulting interference pattern depends on the relative phase, frequency, amplitude, and polarization in the case of vectorial waves such as light. Constructive interference occurs when relative phases enhance the wave amplitudes, while destructive interference results from partial or complete cancellation due to opposing displacements. By contrast, since in quantum mechanics only observable effects carry physical significance, a rigorous description of interference in this context must include the measurement apparatus that specifies exactly what is being observed, as formalized, for instance, in the theory of coherence and quantum photodetection~\cite{Glauber1963, Glauber1963_1, glauber2006, glauber2007quantum}. More recently, further deepening the importance of the observer, interference patterns in double-slit experiments could be reinterpreted in terms of states that couple (bright) and that do not couple (dark) to the detectors, thus revealing new aspects of such fundamental experiments~\cite{PhysRevLett.134.133603}. In this way, it is essential to use the term \textit{interference} with care.


In classical physics, electromagnetic waves (e.g., optical light) and mechanical waves (e.g., sound) do not interfere directly in the classical sense, as they are excitations of different physical substrates: electromagnetic fields in vacuum or dielectric media, and elastic deformations in matter, respectively. Nevertheless, indirect coupling mechanisms can lead to interference-like phenomena. Acoustic waves can modulate the refractive index of a medium, enabling interaction with light through acousto-optic diffraction~\cite{Eggleton2019,Safavi-Naeini_2011}. While this is not classical interference, such coupling yields intensity modulations reminiscent of interference fringes. Similarly, in plasmonic systems~\cite{Tame2013} and optomechanical cavities~\cite{RevModPhys.86.1391, Verhagen2012}, coherent interactions between electromagnetic and mechanical modes can produce interference-like effects, manifesting as controllable transmission or absorption patterns. 

In this work, we adopt a refined quantum-optical perspective, based on a system-observer model, to demonstrate that even waves of distinct physical origin may exhibit well-controlled interference phenomena. By engineering their interactions with the detection apparatus, we show that both perfectly constructive and destructive interference can be realized, thereby generalizing the concept of interference beyond its conventional constraints. 

\section{Wave Interference in Trapped-Ion Systems}

The mechanical properties of quantum matter can often be described by quasiparticles specific to the system under study. In this work, we investigate a system of $N$ identical trapped ions (all of them with mass $m$), where phonons characterize their quantized vibrational modes. The interaction between a given ion in this system and an external electromagnetic field is governed by the dipole Hamiltonian~\cite{Leibfried2003}
\begin{equation}
\label{Eq:dipole}
    \hat{H}_{\rm int}= -\hat{\bm{d}} \cdot \hat{\bm{E}},
\end{equation}
being $\hat{\bm{d}}$ the dipole operator of the electronic system and $\hat{\bm{E}}$ the electric field operator, which enables controlled modifications of both the motional and electronic states of each ion individually, depending on the frequency, phase, and intensity of the driving fields. Additionally, to provide a complete quantum description of the system dynamics, we consider the two-level approximation of the electronic structure of the ions, allowing us to write their Hamiltonian as ($\hbar=1$)
\begin{equation}
    \hat{H}_{\rm el} = \frac{\omega_0}{2}\sum_{n=1}^N  \hat{\sigma}^{(n)}_z,
\end{equation}
\noindent where $\omega_0$ is the transition frequency between the electronic states, assumed to be the same for all ions, and $\hat{\sigma}^{(n)}_z$ is the Pauli-Z operator acting on the $n$-th ion. Now, with the general description for the system, we can deepen the studies into the scope of interest, the observation and control of interference in quantum systems. We start with the single-ion case, where interference between its stationary vibrational mode and an external electromagnetic field can take place. Then, in the following, we analyze the $N$ trapped-ion system, where the interference between a propagating mechanical pulse with an electromagnetic pulse is investigated.

\subsection{Single trapped-ion system}

\begin{figure}
\centering
    \begin{subcaptiongroup}
    \phantomcaption \label{subfig:single_ion}
    \phantomcaption \label{subfig:plot_single}
    \end{subcaptiongroup}
    \includegraphics[width = \columnwidth]{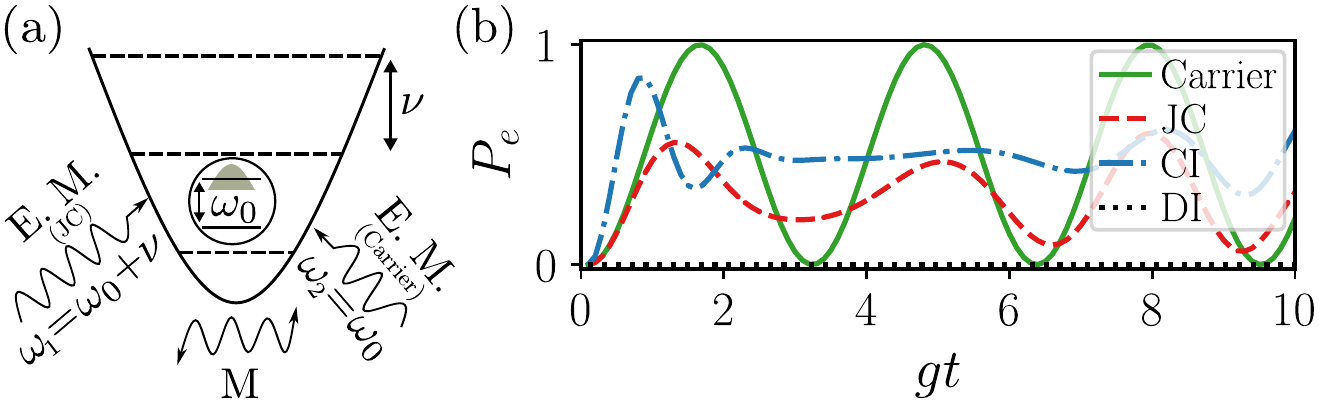}
    \caption{\justifying (\subref{subfig:single_ion}) Single-trapped ion system, characterized by its vibrational ($\nu$) and electronic transition ($\omega_0$) frequencies. It interacts with external electromagnetic (E.M.) fields with frequencies $\omega_1=\omega_0+\nu$ and $\omega_2=\omega_0$, that generate Jaynes-Cummings (JC) and Carrier interactions, respectively. When its vibrational mode (represented by M) is initialized in the coherent state $\vert \alpha \rangle$, with $|\alpha| = \frac{\Omega_2}{\Omega_1\eta} = 1$ (where $\Omega_l$, $l=1,2$, is the respective Rabi frequency of the E.M. fields, and $\eta$ is Lamb-Dicke parameter), the interaction results in constructive (destructive) interference, depending on whether $\alpha$ is positive (negative). (\subref{subfig:plot_single}) Excited state populations of the ion as a function of the normalized time $gt$, without dissipative effects, in four different possible situations: (i) Carrier (dotted line): when only the Carrier interaction is turned on; (ii) JC (dashed dotted-line): when only the JC interaction is turned on; (iii) CI (solid line): when $\alpha = |\alpha|$ and the Carrier interaction amplifies the JC interaction, enhancing the population transfer from the motional mode to the electronic state, \textit{i.e.}, a constructive interference; (iv) DI (dashed line): when $\alpha = -|\alpha|$ and the Carrier interaction nullifies the JC interaction, making the trapped ion transparent to the dipole interaction, \textit{i.e.}, a destructive interference.}
    \label{fig_1}
\end{figure}

We start analyzing the simplest case of a single trapped ion, as depicted in Fig.~\ref{subfig:single_ion}, whose motional Hamiltonian is described by the quantized harmonic oscillator Hamiltonian
\begin{equation}\label{Eq:H_motion_single}
    \hat{H}^{\rm single}_{\rm motion} = \nu \hat{a}^\dagger \hat{a},
\end{equation}
\noindent where $\nu$ is the trap frequency of the vibrational mode, which is described by the phonon creation (annihilation) operator $\hat{a}^\dagger$ $(\hat{a})$ in the Fock basis.

Now, for a laser field characterized by wave vector $\bm{k} = (k_x, k_y, k_z)$, frequency $\omega$, and phase $\varphi$, the interaction Hamiltonian with a single ion takes the form:
\begin{equation}\label{Eq:full_dipole_ham}\small
\hat{H}_{\rm int} = \frac{\Omega(t)}{2} \left( |g\rangle\langle e| + |e\rangle\langle g|\right) \left [ e^{i(\bm{k}\cdot \bm{\hat{r}}-\omega t + \varphi)} + e^{-i(\bm{k}\cdot \bm{\hat{r}}-\omega t + \varphi)}\right],
\end{equation}
\noindent where $\Omega(t)$ is the time-dependent Rabi frequency, $|g\rangle$ ($|e\rangle$) denotes the ground (excited) electronic state, and $\bm{\hat{r}}$ is the position operator corresponding to the ion’s vibrational mode. Considering the ion confined along a single spatial direction $\bm{x}$, we can express the spatial phase factor as $\bm{k}\cdot \bm{\hat{r}} = \eta (\hat{a}+\hat{a}^\dagger)$, where $\eta = k_x \sqrt{\hslash/(2m\nu)}$ is the Lamb-Dicke parameter. Therefore, the full Hamiltonian in the Schrödinger picture reads
\begin{equation}\label{Eq:Hfull}
    \hat{H}_{\rm full} = \hat{H}_{\rm el} + \hat{H}^{\rm single}_{\rm motion} + \hat{H}_{\rm int}.
\end{equation}
\noindent Moving to the interaction picture, through the unitary transformation $e^{-i\hat{H}_0 t/\hbar}$,  with \(\hat{H}_0 = \hat{H}_{\rm el} + \hat{H}^{\rm single}_{\rm motion}\), and applying the rotating wave approximation (RWA),  the time-dependent Hamiltonian becomes
%
%
\begin{multline}\label{interaction_1mode}
    \hat{H}_{I}(t) =  \frac{\hbar\Omega (t)}{2} \left( \hat{\sigma}_+ \, e^{-i(\delta t - \varphi)} e^{i\hat{\gamma}} + \hat{\sigma}_- \, e^{i(\delta t - \varphi)} e^{-i\hat{\gamma}^\dagger} \right),
\end{multline}
\noindent where $\delta = \omega-\omega_0$ is the detuning between the laser and the electronic transition frequencies, and $\hat{\gamma} = \eta(\hat{a}\, e^{-i\nu t}+\hat{a}^\dagger \, e^{i\nu t})$. 

By adjusting the frequency of the laser, different electronic-vibrational interactions can be derived. For instance, the effective Hamiltonian acting over the ion can take the Jaynes-Cummings (JC) interaction form ~\cite{Leibfried2003}
\begin{equation*}
    \hat{H}_{\rm JC} = \frac{g}{2} \hat{a} \hat{\sigma}^+ + \frac{g^*}{2}\hat{a}^\dagger\hat{\sigma}^- ,
\end{equation*}
\noindent with $g=\Omega_1 (t) \eta$, by employing a laser with frequency $\omega = \omega_1$ such that $\delta_1 = \omega_1 - \omega_0 = \nu$, setting $\Omega(t) = \Omega_1(t)$, and applying the RWA in Eq.~\eqref{interaction_1mode}.
This interaction will result in oscillations in the population of electronic states, known as Rabi oscillations~\cite{Xie_2017,PhysRevLett.74.4091,PhysRevLett.75.4714}. 

For the purpose of our work, as will be better motivated in the latter, we can consider the initial vibrational mode state of the ion in the given quasi-classical coherent state 
\begin{equation*}
    |\alpha \rangle = e^{-|\alpha| ^2/2 }\sum_{n=1}^\infty \frac{\alpha^n}{\sqrt{n!}}|n\rangle, 
\end{equation*}
\noindent where $\vert n\rangle$ is the $n$-th Fock state, while its electronic state is initialized in the ground state. For this initial state, the effective JC interaction can be approximated as
\begin{equation}\label{effective_JC}
    \hat{H}^{\rm eff}_{\rm JC} = \frac{\tilde{\Omega}(t)}{2} \hat{\sigma}^+ + \frac{\tilde{\Omega}(t)^*}{2} \hat{\sigma}^-,
\end{equation}
\noindent with $\tilde{\Omega}(t) = \eta \alpha\Omega_1(t)$. 

Going further, if the ion interacts with a second source of light (frequency $\omega_2$) in resonance to the electronic transition, i.e., $\delta_2 =\omega_2-\omega_0= 0$, it is possible to generate, after another RWA, the Carrier interaction, where the Hamiltonian of Eq.~\eqref{interaction_1mode} becomes (with $\Omega(t) = \Omega_2(t)$) 
\begin{equation}\label{H_Carrier}
    \hat{H}_{\rm Carrier} = \frac{\Omega_2(t)}{2} \left( \hat{\sigma}^+ +  \hat{\sigma}^-\right).
\end{equation}
\noindent Thus, by appropriately choosing the initial vibrational state $|\alpha\rangle$, we can control the effective interaction experienced by the ion in both its vibrational and electronic degrees of freedom. Under these conditions, the exchange of energy between the electromagnetic and mechanical waves with the electronic states can be amplified or suppressed depending on the phase of the complex number $\alpha$. In particular, for constant Rabi frequencies $\Omega_1$ and $\Omega_2$, setting $|\alpha| = \Omega_2/\left(\eta\Omega_1\right)$ creates an effect analogous to interference between the vibrational motion of the ion and the applied driving fields. Specifically, this phase-dependent behavior resembles constructive and destructive interference, as the resulting interaction strength changes according to the phase of $\alpha$. For instance, setting $\Omega_2/(\eta\Omega_1) = 1$ allows us to understand different dynamics of the circuit when $|\alpha|=1$. This case, in the absence of dissipative effects, is shown in Fig.~\ref{subfig:plot_single}, where constructive interference enhances the JC interaction, thereby increasing the ion's energy absorption efficiency, whereas destructive interference suppresses the interaction. Equivalently, since in our scheme this relative phase depends explicitly on the ion's vibrational state, we can tune the lasers' Rabi frequencies in order to exploit either effect instead.

The previous result motivates us to examine the possibility of engineering a similar behavior in a larger system, where the coherent state propagates, such as a phonon wave. To do so, we proceed with the analysis of a system of multiple ions, which paves the way to investigate more complex phenomena, such as wave propagation and interference in a scalable quantum medium. Additionally, this system's design has some mechanisms, such as the Rabi frequencies of the two lasers, that are easily managed. This feature will be employed in the next section, where we study interference effects in an ion chain.

\subsection{Trapped-ion chain}

Considering a one‑dimensional chain of $N$ ions, we prepare the vibrational state of the system as $|\alpha\rangle |0\rangle^{ \otimes N}$, where the first ion is initialized in the coherent state $|\alpha\rangle$ and the others in their respective ground states $\ket{0}$. The choice of a coherent state is primarily because it closely mirrors classical wave behavior, possessing a well-defined amplitude and phase. Furthermore, this state is conveniently an eigenstate of the annihilation operator, $\hat{a}|\alpha\rangle = \alpha\vert\alpha\rangle$, whose property is preserved under the coupled harmonic dynamics of the ion chain. Thus, as the coherent wave packet propagates, each local oscillator's mode continues to behave as a coherent state, albeit with a time-dependent amplitude. This persistent coherence character through the chain enables a continuous and coherent mapping of phase and amplitude information along the system, favoring the study of interference effects and the reliable implementation of vibrational-state-dependent interactions, such as those involving the JC and Carrier couplings presented earlier. Hence, the use of coherent states facilitates both physical intuition and theoretical tractability, while ensuring compatibility with the system’s designed interference dynamics.


When we load more than one ion in the same harmonic trapping potential, the Coulombic repulsion between the ions generates a coupling between their vibrational modes. However, as Coulombic interaction is a short-range force, such vibrational coupling becomes restricted to the first neighbors only~\cite{PhysRevLett.120.073001}. On the other hand, for small perturbations around the equilibrium position of each ion, their motions can be appropriately described by harmonic oscillators. In this way, the Hamiltonian that describes the vibrational motion of all ions can be written as a collection of first-neighbor coupled harmonic oscillators, such that $\hat{H}^{\rm chain}_{\rm motion} = \hat{H}_0 + \hat{H}_{NN}$, with $\hat{H}_0=
\sum_{n=1}^N \nu_{n} \hat{a}_{n}^\dagger \hat{a}_{n}$, and  
\begin{align}\label{Eq:H_motion}
    \hat{H}_{NN} =  J \sum_{n=1}^{N-1} (\hat{a}^{\dagger}_n \hat{a}_{n+1} + \hat{a}^{\dagger}_{n+1} \hat{a}_{n}),
\end{align}
\noindent where $\nu_n$ is the frequency of the $n$-th vibrational mode, and $J$ is the coupling strength between neighbors, assumed equal for all ions, for simplicity. 
\begin{figure}[htp!]
    \begin{centering}
    \begin{subcaptiongroup}
    \phantomcaption \label{subfig:chain}
    \phantomcaption \label{subfig:chain_destructive}
    \phantomcaption \label{subfig:plots}
    \end{subcaptiongroup}
    \end{centering}
    \includegraphics[width=\columnwidth]{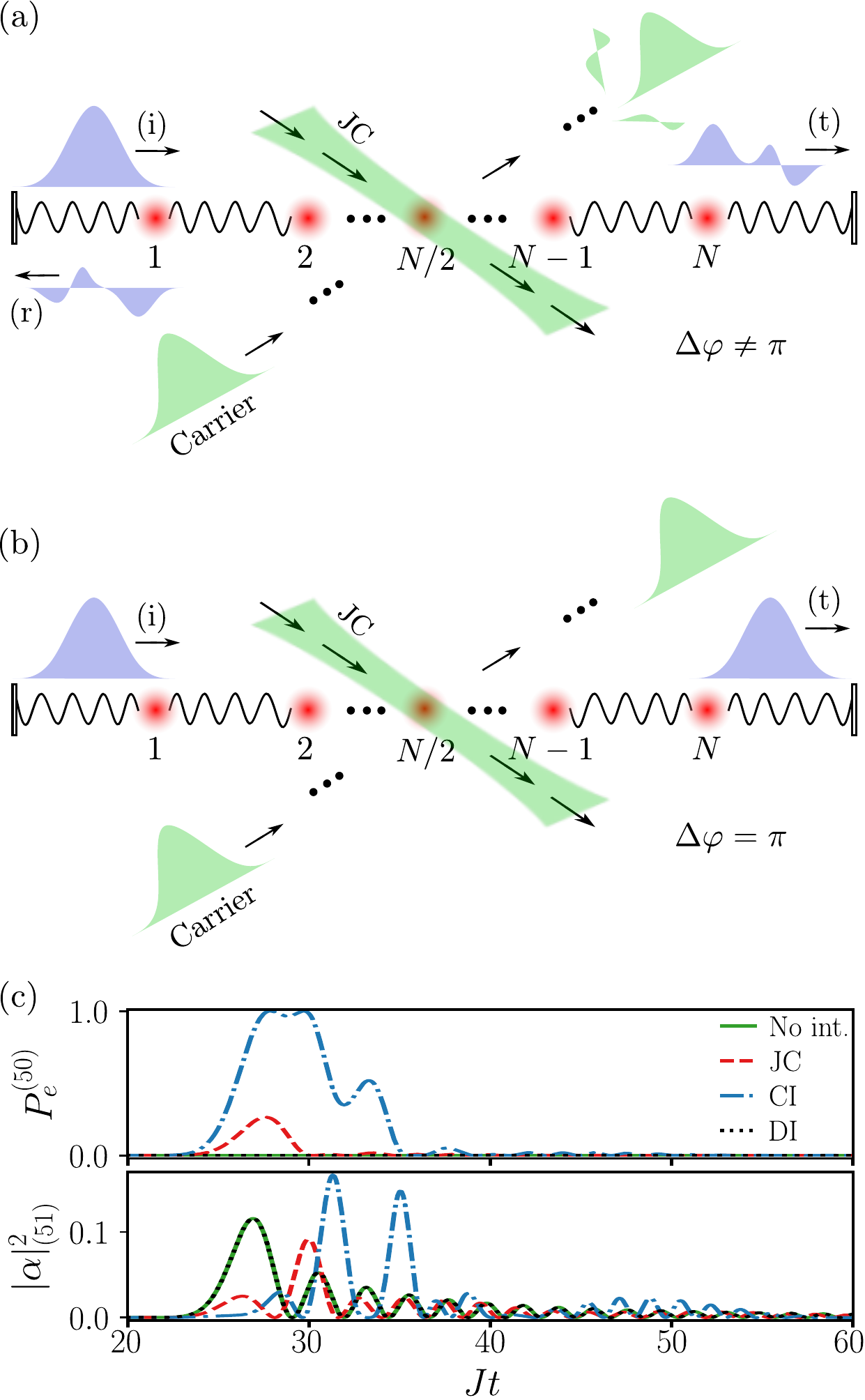}
    
    \caption{\justifying 
    A chain of $N$ trapped ions with two external lasers promoting continuously JC and modulated (through $\Omega_2(t)$) Carrier interactions on the middle ion of the chain, following Eq.~\eqref{Eq:full_ham}. Panel (\subref{subfig:chain}) shows the light and the mechanical wave pulses scattering when $\Delta\varphi \neq \pi$, and panel (\subref{subfig:chain_destructive}) shows the perfect canceling of the interaction with the electronic states of the ion when $\Delta\varphi = \pi$. Hence, adjusting the Rabi frequency $\Omega_2(t)$, we can make the ion transparent to electronic excitations. Considering $g/J = 1$, and the initial state of the first ion $\alpha_1 = 1$,  panel (\subref{subfig:plots}) shows the populations of the excited electronic state of the middle ion $P^{(50)}_e$ and the motional state of the subsequent ion $|\alpha|^2_{(51)}$ as function of the dimensionless time $J t$ for different scenarios: (i) where no laser is applied (No int.), (ii) where only the Jaynes-Cummings driving is applied on the middle ion (JC), (iii) when $\Delta \varphi = 0$ in Eq.~\eqref{Eq:phase_rel}, achieving constructive interaction (CI), and (iv) when $\Delta \varphi = \pi$, achieving destructive interaction (DI).}
    \label{fig_2}
\end{figure}

Now, we consider laser fields continuously applied on the central ion $m= \lfloor (N+1)/2 \rfloor$, to generate the Jaynes-Cummings and Carrier interactions: a first laser with constant Rabi frequency $\Omega_1$, that produces the interaction Hamiltonian $H_{\rm JC}^{(m)} = g/2 \left(\hat{a}_{m} \hat \sigma_m^+ + \hat{a}_m ^\dagger\hat \sigma_m ^-\right)$, and a second laser generating a modulated Carrier interaction with Rabi frequency $\Omega_2(t)$, given by $ H^{(m)}_{\rm Carrier}=\frac{1}{2}\left(\Omega_2(t) \sigma_{(m)}^+ + \Omega_2^*(t)\sigma_{(m)}^-\right)$. Thus, the full Hamiltonian becomes
\begin{align}
\label{Eq:full_ham}
    \hat{H} &= \hat{H}_{0} + \hat{H}_{NN} + \hat H_{\rm JC}^{(m)} + \hat H^{(m)}_{\rm Carrier}. 
\end{align}

In order to study the dynamics ruled by this Hamiltonian accurately and ensure negligible boundary effects, we simulate chains with a large number of ions, specifically $N=100$~\cite{shapira2023fastdesignscalingmultiqubit} (resulting in $m=50$), in order to suppress information fallback. This phenomenon is observed in finite systems, where the wave packet reaching the end of the chain reflects back and superposes with itself. By working with a sufficiently large system, such reflections occur only at later times, outside the window of interest, thereby preserving the integrity of the simulated coherent transport. In addition, for simplicity, we have assumed the same vibrational frequency for all ions, i.e., $\nu_m = \nu$, the coherent initial state of the first ion is set $\vert \alpha =1 \rangle$, and that there are no dissipation effects. To simulate such a system, we work in the interaction picture and the semiclassical approximation is applied, in which the vibrational modes are approximated to time-dependent classical coherent amplitudes $\alpha$ and the expected values of the Pauli operators approximate the time evolution of electronic operators (see Appendix~\ref{Appendix1} for further details).

To a better comprehension of how the dynamics evolve and the information is transmitted across the chain, consider that the modulated Rabi frequency $\Omega_2(t)$ from the Carrier interaction and the instantaneous coherent state amplitude of the middle ion, $\alpha_m(t)$ (in the absence of external fields), satisfy $\Omega_2(t) = g\alpha_m(t)e^{i \Delta\varphi}$, resulting in
\begin{equation}\small
    \label{Eq:phase_rel}
    \Omega_2(t)+g\alpha_m(t) = g\alpha_m (t)\left(e^{i \Delta\varphi} + 1\right).
\end{equation}
\noindent where $\Delta\varphi$ is a criterion phase that helps to analyze the two different regimes of the system, pictorially presented in Figs.~\ref{subfig:chain} and~\ref{subfig:chain_destructive}. From the above equation, when $\Delta\varphi \neq \pi$, i.e., when the Rabi frequency of the carrier pulse does not perfectly counter the JC interaction, with effective coupling $g\alpha_m(t)$, the combined waves excite the electronic state of the ion $m$. In fact, for $\Delta\varphi=2l\pi$ ($l=0,1,2,...$), the resulting interaction becomes twice as strong as that of the individual interacting waves, characterizing constructive interference. Conversely, when $\Delta\varphi$ is any odd multiple of $\pi$, the Carrier interaction rightly balances out the energy exchange of the JC Hamiltonian and no excitation of the electronic levels happens, as it happens in usual destructive interference.


Deepening into the analysis of both cases, we elaborated Fig.~\ref{subfig:plots}, where we show the populations of the excited electronic state of the middle ion $P^{(50)}_e$ and the square of the motional state amplitude of the subsequent ion $|\alpha|^2
_{(51)}$. In a preliminary case, when no interaction due to external fields is occurring -- green solid lines, we can notice that all the energy arriving at the $m$-th ion is coherently transmitted to the following one, and the propagation remains unaffected, as it would be expected. However, once the first (and only, for now) laser is applied, the presence of the constant JC coupling with the electronic state of the $m$-th ion modifies the energy transfer across the chain -- red dashed lines. Consequently, with the JC interaction turned on, a fraction of the energy is now allowed to occupy the electronic state of the middle ion, reducing the amount of energy that the transmitted wave packet has. This behavior is analogous to the reflection and transmission of light at the boundary between two optical media. In this analogy, the coupling strength $g$ plays a role similar to that of the refractive index, which determines the impedance mismatch for phonon propagation~\cite{10.1063/1.4893789}. In this sense, a larger coupling acts as a high refractive index contrast, leading to complete reflection, which would be characterized by the confinement of phonons between the first $m$ ions of the chain. Contrarily, smaller couplings allow partial transmission of the excitation, akin to light refraction across an interface.


Now, when the second laser is also driven onto the middle ion, both interactions are allowed to occur simultaneously, and then, one of the previous regimes will be observed. When we have $\Delta\varphi \neq \pi$, the two amplitudes do not cancel out, resulting in the partial excitation of the electronic state of the $m$-th ion. More specifically, if $\Delta\varphi = 0$, constructive interference is created -- blue dashed-dotted lines -- and non-negligible information is continuously exchanged between the vibrational mode of the ion and its electronic state through the JC interaction. Thus, this energy is scattered via neighbor coupling, with the transmitted pulse having a smaller energy and a different coherent shape.  On the other hand, when $\Delta\varphi = \pi$, the Carrier interaction strength is exactly the opposite to the effective excitation promoted in the electronic state by the JC interaction, and the destructive interference is arranged -- black dotted lines. Consequently, the electronic state remains unpopulated, and the wave packet flows along the chain without disturbances.


The interference between the Carrier driving (electromagnetic wave) and the vibrational pulse (mechanical wave), mediated through the electronic state of the $m$-th ion, can already be anticipated from the single-trapped ion results. By examining Eq.~\eqref{Eq:phase_rel}, the potential for such interference is immediately evident, particularly highlighting the state-dependence of this effect. Therefore, precisely adjusting the Carrier Rabi frequency $\Omega_2(t)$, while properly accounting for the JC coupling strength and the input coherent state, allows accurate information flow control across the ion chain.

An interesting application can be achieved when the system is in the phonon blockade regime~\cite{Abo2022}, i.e., $g \gg J$. Under this condition, when carefully selecting the modulation of the Carrier Rabi frequency, a selective filter is engineered, actively controlling the phononic transmission through the ion chain. Such selectivity provides functionality analogous to that of a quantum transistor or highly efficient switches~\cite{alan_transistor}, allowing or not the propagation of information. More remarkably, the tunability of the Carrier Rabi frequency manipulates the wave packets based on their amplitude and phase characteristics, which enables precise control of the transmission rate of specific states, while providing for enhancements or nullification of electronic excitations.

In usual trapped ion experiments, the Rabi frequency $\Omega_1$ of the JC driving can reach up to $0.8 \times 2\pi\,\text{MHz}$~\cite{chen2024studydecoherenceradiallocal,löschnauer2024scalablehighfidelityallelectroniccontrol} while Lamb-Dicke parameters are around $\eta \lesssim 0.2$~\cite{PhysRevLett.110.263002,PhysRevResearch.5.023022}. Also, considering an initial coherent state with a mean number equal to unity, the effective JC coupling can be up to $g=160 \times 2\pi \text{ kHz}$, which is much higher than the hopping rate $J$ between the vibrational modes, typically around $3\times2\pi$~KHz~\cite{PhysRevLett.120.073001}. In this sense, the feasibility for regimes where $g \gg J$ is within reach, experimentally allowing phonon blockade. In addition, when the Carrier interaction is turned on, whose Rabi frequency $\Omega_2(t)$ can be modulated to have the same order of magnitude as $g$, the interference effects presented earlier are achieved. These features establish a clear pathway toward assessing the proposed device. Furthermore, due to the state dependence of the effective JC coupling, the precise control of the modulated Carrier Rabi frequency enables the engineering of a coherent-state selective filter, which works as a wave packet quantum transistor while allowing for precise manipulation of phononic transmission.


\section{Conclusions}\label{Conclusions}


In this work, we extend the concept of interference to waves of different natures, a concept absent in classical theory, i.e., we show that an interference phenomenon between electromagnetic and mechanical waves can take place. To observe this effect, we consider the interaction of those waves with the electronic levels of a trapped ion: By modulating light-matter interactions and considering the propagation of information through a single trapped ion or a mechanical chain, we enable the interference between pulses of light and mechanical waves, resulting in either constructive or destructive interference depending on their phase relationship. Since the ion's electronic state can be coupled to vibrational modes using light, the combined use of JC and Carrier interactions with the same ion can either enhance or perfectly cancel the wave packet propagating through the chain.


Considering the $N$-ion chain, with first neighbor vibrational couplings, when the JC coupling is much higher than the mechanical coupling, the driven qubit confines the dynamics to the first half of the vibrational modes of the chain, which we mentioned as the phonon blockade effect~\cite{Abo2022}, a phenomenon similar to the photon blockade~\cite{PhysRevLett.79.1467}. When the modulated Carrier is also applied to the same ion, phonon transmission can be controlled by tuning the Rabi frequency, working as a transistor that depends on the incoming coherent pulse, effectively selecting and filtering the information transmitted. In addition, the ion chain employed here, with adjustable mechanical and electromagnetic couplings, could be used to simulate the optical properties of media, allowing investigations, e.g., on how to control the transmission, refraction, and diffraction of acoustic or electromagnetic waves by external light or mechanical forces.

Looking forward, these hybrid interference processes open a versatile route toward quantum-controlled signal processing. The same mechanism could be exploited to build reconfigurable phonon routers~\cite{PhysRevLett.134.210601,Habraken_2012}, vibration-noise cancellers, or phase-sensitive acoustic sensors. Extending the concept to other platforms, such as optomechanical cavities, surface-acoustic-wave circuits, and plasmonic hybrids, could yield integrated components for quantum communication, precision metrology, and chip-scale information processing where electromagnetic and mechanical degrees of freedom are harnessed cooperatively rather than treated as separate domains~\cite{10.3389/frqst.2022.1091691,Jiang2016,Meher2022,10.1063/10.0028127,PRXQuantum.5.010101}. Finally, the semiclassical approach employed here is already enough to introduce our controllable mechanical-electromagnetic wave interference. However, further investigation would require a completely quantum model to allow one to analyze the dynamical behavior of the system with other photonic-phononic states, like general collective bright and dark states~\cite{PhysRevLett.134.133603} of modes of different nature.

\section{Acknowledgments}
This work was supported by the Coordenaç\~{a}o de Aperfeiçoamento de Pessoal de  N\'{i}vel Superior (CAPES) - Finance Code 001 and S\~{a}o Paulo Research Foundation (FAPESP) grants No. 2022/00209-6 and 2022/10218-2. C.J.V.-B. is also grateful for the support of the National Council for Scientific and Technological Development (CNPq) Grant No. 311612/2021-0. This work is also part of the CNPq Grant No. 140467/2022-0.

\bibliography{refs1}

\appendix

\section{Semiclassical simulation of the quantum system}\label{Appendix1}

The quantum dynamics were solved using a semiclassical approximation to the Hamiltonian given in Eq.~\eqref{Eq:full_ham}, working in the interaction picture. This approach treats the vibrational modes as time-dependent classical coherent amplitudes $\alpha$ and the expected values of the Pauli operators as a description for the internal states of the ions. The resulting set of coupled differential equations governs the time evolution of both the motional and electronic degrees of freedom. Explicitly, the system is described by:
\begin{equation*}
    \frac{d \alpha_k ^R}{dt} =  J\left(\alpha_{k+1}^I + \alpha_{k-1}^I\right) - \delta_{k,N/2}\, \frac{g}{2}\sigma^y_{N/2},
\end{equation*}
\begin{equation*}
    \frac{d \alpha_k ^I}{dt} = -  J\left(\alpha_{k+1}^R + \alpha_{k-1}^R\right) - \delta_{k,N/2} \, \frac{g}{2}\sigma^x_{N/2},
\end{equation*}
\begin{equation*}
    \frac{d\sigma_k^x}{dt} =- \delta_{k,N/2} \, 2g\alpha_{N/2}^I \sigma_{N/2}^z - \delta_{k,N/2} \, 2\text{Im}(\Omega)\sigma_{N/2}^z,
\end{equation*}
\begin{equation*}
    \frac{d\sigma_k^y}{dt} = - \delta_{k,N/2} \, 2g\alpha_{N/2}^R \sigma_{N/2}^z -\delta_{k,N/2} \,  2\text{Re}(\Omega)\sigma_{N/2}^z,
\end{equation*}
\begin{multline*}
    \frac{d\sigma_k^z}{dt} = -\delta_{k,N/2} \, 2g\sigma^x_{N/2} \alpha^I_{N/2} + \delta_{k,N/2} \, 2g\sigma_{N/2}^y \alpha^R_{N/2} - \\  -\delta_{k,N/2} \, 2\text{Im}(\Omega) \sigma_{N/2}^x + \delta_{k,N/2} \, 2 \text{Re}(\Omega) \sigma_{N/2}^y,
\end{multline*}
\noindent where $\alpha_k^R$ and $\alpha_k^I$ are the real and imaginary parts of the semiclassical amplitude $\alpha_k(t)$, representing the motional (phonon) degree of freedom at site $k$, and $\sigma_k^{x,y,z}$ are the semiclassical expected values of the usual Pauli operators for the $k$-th qubit. The Kronecker delta function $\delta_{i,j}$ ensures that both the Jaynes–Cummings and Carrier interactions are applied only at the middle ion, $k = N/2$.

The resulting set of coupled differential equations is stiff due to the presence of vastly different timescales in the system—namely, fast oscillations from the Carrier frequency combined with the slower dynamics of phonon propagation and electronic state transitions. To accurately and efficiently integrate these equations over long simulation times, we employ the Explicit Singly Diagonally Implicit Runge–Kutta (ESDIRK) method of 5th order with seven stages, as implemented in the DifferentialEquations.jl package \cite{rackauckas2017differentialequations} in Julia \cite{bezanson2017julia}. This method is well-suited for stiff systems because it provides excellent stability properties, particularly for oscillatory problems like ours, while maintaining high-order accuracy. Moreover, the use of implicit stages helps suppress numerical instabilities that typically arise in high-frequency regimes or when modeling systems with sharp transitions, such as sudden changes in the Carrier interaction phase.

This semiclassical treatment is valid in regimes where the phonon amplitudes remain significant and coherent and allows for efficient numerical integration over long timescales. Nevertheless, a complete quantum treatment would be necessary for systems with substantial quantum correlations or entanglement between phonons and qubits. The comparison between semiclassical and fully quantum models remains an open direction for quantifying nonclassical effects in coherent phonon transport. Unlike a full quantum simulation that tracks entanglement and quantum fluctuations, this approach aims to model the coherent dynamics efficiently while ignoring quantum correlations.

\end{document}